\documentclass[aps,prb,twocolumn,showpacs,superscriptaddress,groupedaddress,floatfix,preprintnumbers]{revtex4-1}
\usepackage[pdftex]{graphicx} 
\usepackage{dcolumn}   
\usepackage{bm}        
\usepackage{amssymb}   
\usepackage{amsmath}   
\usepackage{natbib}

\PassOptionsToPackage{caption=false}{subfig}
\usepackage{subfig}

\newcommand{\beq}{\begin{equation}}
\newcommand{\eeq}{\end{equation}}

\newcommand{\beqa}{\begin{eqnarray}}
\newcommand{\eeqa}{\end{eqnarray}}

\newcommand \nline {\nonumber \\}

\newcommand \freedm {{\it F}}
\newcommand \pxpy[2] {\frac{\partial #1}{\partial #2}}

\begin{document}

\title{Phase field crystal modeling of early stage precipitation and clustering}
\author{$^{1,2}$Vahid Fallah, $^3$Jonathan Stolle, $^1$Nana Ofori-Opoku, $^2$Shahrzad Esmaeili, $^{1,4}$Nikolas Provatas}
\affiliation{$^1$Department of Materials Science and Engineering, 
McMaster University,1280 Main Street West, Hamilton, Canada L8S-4L7}
\affiliation{$^2$Mechanical and Mechatronics Engineering Department, 
University of Waterloo,200 University Avenue West, Waterloo, Canada N2L-3G1}
\affiliation{$^3$Department of Physics and Astronomy, 
McMaster University,1280 Main Street West, Hamilton, Canada L8S-4L7}
\affiliation{$^4$Department of Physics and Centre for the Physics of Materials, 
McGill University, 3600 University Street, Montreal, Canada H3A-2T8}

\begin{abstract}
A phase field crystal model is used to investigate the mechanisms of formation and growth of early clusters in quenched/aged dilute binary alloys, a phenomenon typically outside the scope of molecular dynamics time scales. We show that formation of early sub-critical clusters is triggered by the stress relaxation effect of quenched-in defects, such as dislocations, on the energy barrier and the critical size for nucleation. In particular, through analysis of system energetics, we demonstrate that the growth of sub-critical clusters into overcritical sizes occurs due to the fact that highly strained areas in the lattice locally reduce or even eliminate the free energy barrier for a first-order transition.
\end{abstract}

\maketitle

\section{Introduction}
Clustering in metal alloys is known as the very early stage of first-order transformations within a bulk crystal, which largely influences the mechanical properties of quenched/aged materials. Coherent with the matrix, small clusters of solute atoms have a significantly lower nucleation barrier than the terminal second-phase of a completely different crystal structure. Early clusters nucleate and grow in size forming the so-called GP-zones known to be a metastable precursor of the equilibrium phase~\cite{christian}. The formation and growth mechanisms of the early clusters are poorly understood presently due to the lack of direct atomistic observations of structural changes during the transition process. However, inspired by the observations made on the quenched structures using transmission electron microscopy (TEM)~\cite{marceau10,marceau10-2,nutyen67,ozawa70}, 3D atom probe~\cite{sato04,esmaeili07,marceau10,marceau10-2,biswas11}, and positron annihilation~\cite{somoza02,somoza10,marceau10-2} techniques, the formation of early clusters has been empirically associated with so-called quenched-in defects. Formed within the bulk crystal upon quenching, excess vacancies and/or dislocations loops have been presumed to decrease the energy barrier for nucleation facilitating cluster formation~\cite{esmaeili07,ozawa70,somoza10,marceau10,marceau10-2}.

Understanding solute clustering mechanisms is of crucial importance to design an effective age-hardening process producing desired mechanical properties in alloys~\cite{christian}. To our knowledge, no systematic study of the clustering mechanisms has been done using atomic-scale simulation methods such as molecular dynamics (MD) and Monte Carlo (MC) simulations, due to their main restrictions of accessing the relevant time scales of diffusional transformations. Dynamical calculations using classical density functional theory (CDFT) are also inefficient due to the high spatial resolution required to resolve the sharp density spikes in solid phases~\cite{jaatinen09}.

The recently developed phase field crystal (PFC) method~\cite{elder04,elder07,wu10,greenwood10,greenwood11} has shown promise for simulating structural transformations on diffusive time scales. This new formalism carries the essential physics of CDFT without the need to resolve the sharp atomic density peaks. In the most recent PFC formalism developed by Greenwood et al.~\cite{greenwood10,greenwood11,greenwood11-2}, various crystal symmetries can be easily stabilized by construction of relevant correlation kernels. This approach preserves the numerical efficiency of the original PFC model and is able to dynamically simulate the precipitation of solid phases within a parent phase of different crystal symmetry~\cite{greenwood10} and/or chemical composition~\cite{greenwood11-2}.

This letter proposes a new approach to study the clustering phenomenon that relies on atomic-scale simulations using the previously developed alloy PFC model of ref.~\cite{greenwood11-2}. We explore the formation and growth mechanisms of early clusters in a quenched bulk lattice of a supersaturated Al-Cu alloy initially containing quenched-in defects such as dislocations.

\section{Model Structure}
We start with the free energy functional in the binary PFC model~\cite{greenwood11-2},
\begin{align}
\frac{\Delta\freedm}{kT\rho^o} = \int f dr &=
\int \bigg\{\frac {n^2}{2}-\eta\frac {n^3}{6}+\chi\frac {n^4}{12}+\nline(n+1)\Delta\freedm _{mix} &-\frac 1{2}n \int dr'C^n_{eff}(|r-r'|)n'+\alpha|\vec{\nabla}c|^2\bigg\} dr
\label{PFCalEnrgy2}
\end{align}
where $n$ and $c$ represent reduced dimensionless atomic number density and solute concentration fields, respectively. $\eta$ and $\chi$ are coefficients added to fit the ideal energy to a polynomial expansion ($\eta=\chi=1$ describes a Taylor series expansion of the bulk free energy around the reference density) and
\begin{align}
\Delta\freedm_{mix}=\omega\{c\ln(\frac{c}{c_o})+(1-c)\ln(\frac{1-c}{1-c_o})\}
\label{Fmix}
\end{align}
represents the energy density associated with the entropy of mixing. The coefficient $\omega$ is introduced to fit the entropic energy away from the reference composition $c_0$. The parameter $\alpha$ is a coefficient (taken as 1 in this study). These parameters are discussed further in ref.~\cite{greenwood11-2}.

For a binary alloy, Greenwood et al.~\cite{greenwood11-2} introduced the correlation function
\begin{align}
C^n_{eff}=X_1(c)C^{AA}_2 + X_2(c)C^{BB}_2
\label{CorrEff}
\end{align}
, where $X_1(c)=1-3c^2+2c^3$ and $X_2(c)=1-3(1-c)^2+2(1-c)^3$. $C^{AA}_2$ and $C^{BB}_2$ are correlation functions representing, respectively, contributions to the excess free energy for the situations where A atoms are in the preferred crystalline network of B atoms and B atoms which are in a structure preferred by A atoms. The correlation functions $\hat{C}^{ii}_2(\vec{k})$ are defined to have reciprocal space peaks (i.e. $k_j$, corresponding to the inverse of interplanar spacings) determined by the main families of planes in the equilibrium crystal unit cell structure for the $i^{th}$ component. Each peak is represented by the following Gaussian form of width $\alpha_j$, modulated for temperature $σ$ by a Debye-Waller prefactor which accounts for an effective transition temperature $\sigma_{Mj}$~\cite{greenwood11-2}.
\begin{align}
\hat{C}^{ii}_{2j}=e^{-\frac{\sigma^2}{\sigma^2_{Mj}}}e^{-\frac{(k-k_j)^2}{2\alpha^2_j}} 
\label{CorrF}
\end{align}

The equations of motion of the total density and concentration fields follow dissipative dynamics~\cite{archer05}. The total mass density and total reference density per unit volume are defined as $\rho=\rho_A+\rho_B$ and $\rho^o=\rho_A^o+\rho_B^o$, respectively. Thus, the equations of motion can be written for $n$($=\rho/\rho^o-1$) and $c$($=\rho_B/\rho$) as $\pxpy{n}{t}=\vec{\nabla}.\{M_n\vec\nabla(\frac{\delta \Delta F}{\delta n})\}+\eta_n(\sigma,t)$ and $\pxpy{c}{t}=\vec{\nabla}.\{M_c\vec\nabla(\frac{\delta \Delta F}{\delta c})\}+\eta_c(\sigma,t)$, respectively~\cite{greenwood11-2}. $M_n$ and $M_c$ are dimensionless kinetic mobility parameters (equal to 1 in this study). $\eta_n(\sigma,t)$ and $\eta_c(\sigma,t)$ are stochastic noise variables subsuming the role of fast atomic vibrations in density and concentration fields, respectively.

\section{Results}
\subsection{Phase diagram reconstruction}
To examine the equilibrium properties of this binary PFC model for a 2D Al-Cu system, we construct the phase diagram for the coexistence of two square phases. The coexistence lines between the respective phases are obtained by a common tangent construction of the free energy curves of solid and liquid at the reference density ($\bar{n}=0$). Following Greenwood et al.~\cite{greenwood11-2}, the free energy curves of the square phases are calculated using the two-mode approximation of the density fields which is defined by 
\begin{align}
n_i(\vec{r})=\sum_{j=1}^{N_i}A_j\sum_{l=1}^{N_j}e^{2\pi\mathbf{i}\vec{k}_{l,j}.\vec{r}/a_i}
\label{Density}
\end{align}
, where the subscript $i$ denotes a particular solid phase with a lattice spacing $a_i$, and the index $j$ counts the $N_i$ modes of the $i$-phase. $A_j$ is the amplitude of mode $j$ and $l$ is the index over the group of reciprocal space peaks corresponding to mode $j$, $N_j$. Accordingly, $\vec{k}_{l,j}$ is the reciprocal lattice vector normalized to a lattice spacing of 1, corresponding to each index $l$ in the family $j$. The free energy curve for each phase can be calculated as a function of the composition $c$ by substituting the above density field approximation into Eq.~(\ref{PFCalEnrgy2}) and integrating over the unit cell. The resulting crystal free energy is then minimized for the amplitudes $A_j$. For the liquid phase, the amplitude $A_j$ is set to zero and the density is considered as the reference density ($\bar{n}=0$). A more detailed description of this methodology is provided in ref.~\cite{greenwood11-2}.

\begin{figure}[htbp]
\resizebox{3.3in}{!}{\includegraphics{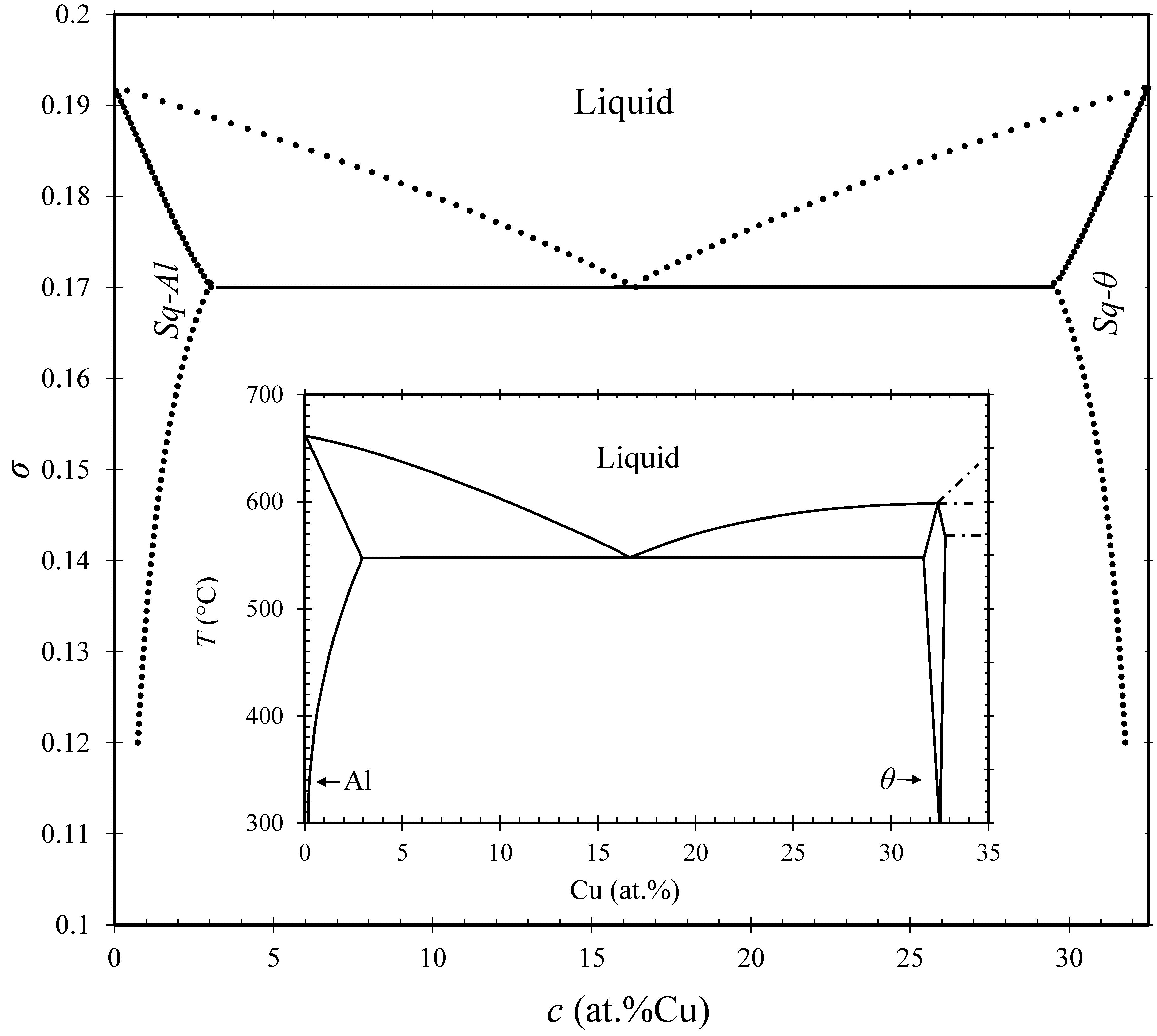}}
\caption{The constructed phase diagram for a square-square system with the inset showing the Al-rich side of the experimental phase diagram of Al-Cu system taken from Ref.~\cite{baker}. The parameters for ideal free energy contribution were $\eta=1.4$ and $\chi=1$, while $\omega=0.005$ and $c_0=0.5$ for entropy of mixing. Widths of the correlations peaks are $\alpha_{11Al}=2.4$, $\alpha_{10Al}=\sqrt{2}\alpha_{11Al}$ (the required ratio to introduce isotropic elastic constants in an square phase~\cite{greenwood11-2}), $\alpha_{11\theta}=2.4$ and $\alpha_{10\theta}=\sqrt{2}\alpha_{11\theta}$. The peak positions for pure Al correspond to $k_{11Al}=2\pi$, $k_{10Al}=\sqrt{2}k_{11Al}$, $k_{11\theta}=(81/38)\pi$ and $k_{10\theta}=\sqrt{2}k_{11\theta}$. The effective transition temperatures are set to $\sigma_{M11Al}=0.55$, $\sigma_{M10Al}=0.55$, $\sigma_{M11\theta}=0.55$ and $\sigma_{M10\theta}=0.55$; The concentration $c$ is rescaled considering the Cu-content in the $\theta$-phase.}
\label{fig:PhaseDiagram}
\end{figure}

In the Al-rich side of the experimental Al-Cu phase diagram, shown in the inset of Fig.~\ref{fig:PhaseDiagram}, there is a eutectic transition between the Al-rich $\alpha$-fcc phase and an intermediate phase $\theta$ (containing $\approx 32.5at.\%$ Cu) with a tetragonal crystal structure. For 2D simulations, in order to approximate these equilibrium properties, we reconstruct the binary phase diagram of Al and $\theta$, both with a square symmetry but differing in Cu-content. The lattice constant (and thus the reciprocal space peaks) of $\theta$ is approximated by interpolating between those of Pure Al and Cu. The solid phase free energy is calculated with a variable lattice constant weighted by concentration $c$ using the interpolation functions $X_1$ and $X_2$. The polynomial fitting parameters in Eq.~(\ref{PFCalEnrgy2}) (namely $\eta$, $\chi$ and $\omega$) and width of various peaks ($\alpha_j$) in the correlation kernel $\hat{C}^{ii}_{2j}$ are then chosen so as to obtain the same compositions for $\alpha$-phase solubility limit and eutectic point as those in the experimental phase diagram. 

\subsection{Simulation of clustering}
With the equilibrium properties obtained above, simulations of clustering were performed on a rectangular mesh with grid spacing $dx=0.125$ and time step $dt=1$. Considering the lattice parameter of 1, each atomic spacing is resolved by 8 mesh spacings. The dynamical equations were solved semi-implicitly in Fourier space for higher efficiency. The initial conditions were chosen to study the proposed dominant role of quenched-in dislocation-type defects in the bulk crystal during the early stage precipitation in dilute Al-Cu alloys quenched from a solutionizing temperature~\cite{nutyen67,ozawa70,somoza02,somoza10,desorbo58}. According to this hypothesis, dislocation loops, generated by excess vacancies, are responsible for local lattice distortions facilitating segregation and diffusion of Cu-atoms, while also driving the system towards a more thermodynamically-stable state~\cite{nutyen67,ozawa70,somoza10,marceau10,marceau10-2}. Therefore, as initial conditions, we use a crystal lattice of uniform composition distorted by introducing dislocations.

\begin{figure}[htbp]
\resizebox{3.3in}{!}{\includegraphics{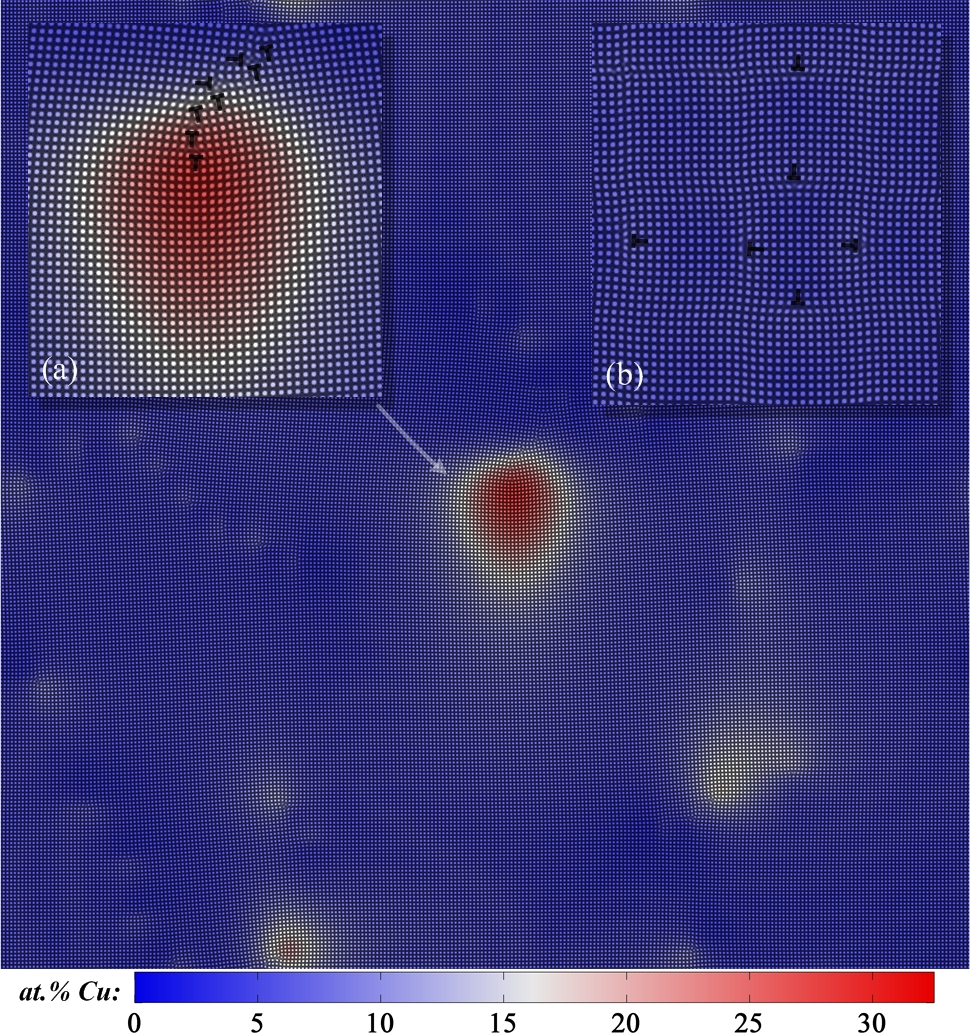}}
\caption{(colour online) PFC simulation of clustering phenomena on a system of 256$\times$256 atoms after 225,000 time steps containing clusters with various sizes and concentrations; (a) The developed structure of a long-lived cluster; (b) The initially distorted structure; For graphical illustration, the concentration field is superimposed on the density field, and ranges from dark blue to dark red as the Cu-content increases.}
\label{fig:Clustering}
\end{figure}

PFC simulation is performed for quench/aging of Al-2at.$\%$Cu from the solutionizing temperature of $\sigma=0.17$ to $\sigma=0.04$ with the initial conditions shown in Fig.~\ref{fig:Clustering}(b). During the simulation, first, small clusters form with a slightly higher Cu-content than that of the matrix. As time progresses, some of these clusters shrink in size and concentration and a few get stabilized (e.g. the cluster shown in Fig.~\ref{fig:Clustering}(a)). In contrast, as expected, quenching the same initial structure from the solutionizing temperature of $\sigma=0.17$ to a temperature within the single-phase $Sq$-$Al$ region, i.e., $\sigma=0.16$, leads to complete removal of distortion. 

\section{Discussion}

\subsection{Evolution of clusters}
The dislocation-induced cluster structure shown in Fig.~\ref{fig:Clustering}(a) is consistent with TEM observations in Al-1.7at.$\%$Cu~\cite{nutyen67} and Al-1.1at.$\%$Cu-0.5at.$\%$Mg alloys~\cite{marceau10,marceau10-2}, where dislocation loops appear in the bulk lattice of the quenched structures. Using resistometric measurements and TEM techniques for Al-1.2at.$\%$Si alloy, Ozawa and Kimura~\cite{ozawa70,ozawa71} have associated the formation of dislocation (or vacancy) loops upon quenching to the coalescence of excess vacancies. They have further suggested that the solute atoms segregate towards the loops stabilizing them into solute clusters. Also, tracing vacancy clusters by positron annihilation, Somoza et al.~\cite{somoza02,somoza10} have proposed that vacancy-Cu pairs are present at the quenched-state in Al-1.74at.$\%$Cu alloy. To our knowledge, our PFC simulations are the first atomic-scale simulations to support the above hypothesis of vacancy/dislocation-mediated solute clustering and nucleation mechanisms of early stage precipitation.

\subsection{Analysis of work of formation}
We further investigated the above mechanisms of cluster formation and growth by analyzing the system energetics for a long-lived cluster. To avoid possible finite size effects, a test with same conditions as those of the above simulation was performed on a larger system, e.g., 512$\times$512 atoms. The strain field caused by the dislocations displacement fields is evaluated by
\begin{align}
\epsilon = \sum_{i=1}^{N_{tri}} \sum_{j=1}^{3}\bigg(\frac{a_{ij}-a_o}{a_o}\bigg)
\label{Strain}
\end{align}
, which is calculated over triangulated density peaks using the Delaunay Triangulation method. $N_{tri}$ is the number of triangles in the field, $a_o$ is the dimensionless equilibrium lattice parameter (the number of grid points resolving one lattice spacing, i.e., 8), $a_{ij}$ is the length of the $j^{th}$ side of the $i^{th}$ triangle. Small clusters, each accompanied by at least one dislocation, appear to be in local equilibrium with the matrix shown in Fig.~\ref{fig:StrainField}(a)). During the simulation, following Fig.~\ref{fig:StrainField}(b) and (c), cluster ``a" continues to grow while, simultaneously, its accompanying dislocation climbs up towards nearby dislocations, creating larger local strain fields (i.e. 0.001, 0.0016 and 0.014 for cluster ``a" in Fig.~\ref{fig:StrainField}(a), (b) and (c), respectively). This mechanism of stress relaxation through solute segregation has been shown through phase-field studies by Leonard and Haataja~\cite{leonard05} to be the main cause of alloy destabilization by structural spinodal decomposition in the presence of dislocations. Also, PFC studies of thin layers deposition by Muralidharan and Haataja~\cite{muralidharan10} indicated that, due to the above mechanism, some immiscible alloys exhibit miscibility gap around the inter-layer interface in the presence of coherency stresses.

\begin{figure*}[htbp]
\resizebox{5.1in}{!}{\includegraphics{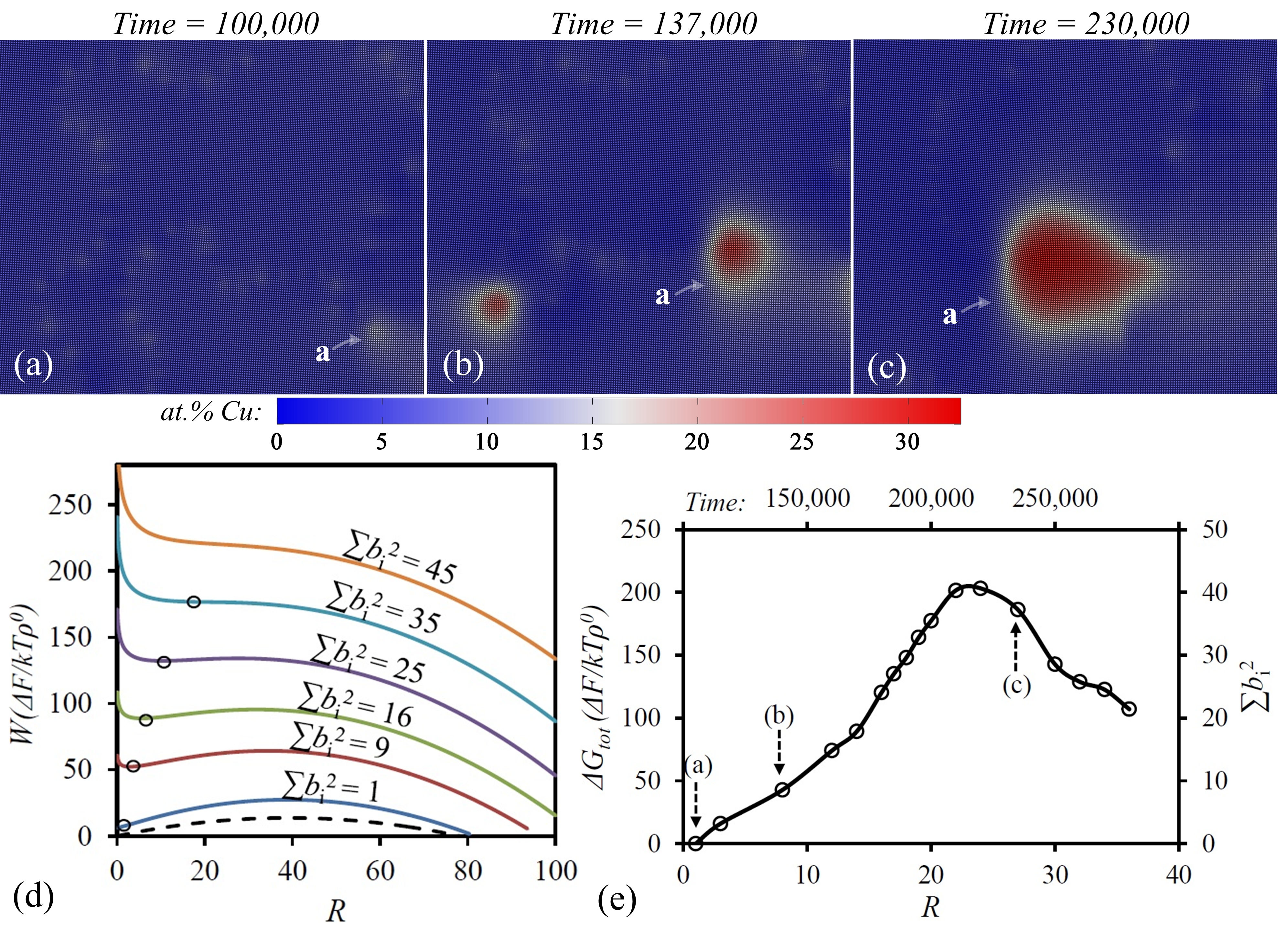}}
\caption{(colour online) (a-c) Snapshots taken at 3 different times showing the structural changes during formation of cluster ``a"; (d) work of formation (evaluated from Eq.(~\ref{Nucl.En})) vs. $R$ for increasing dislocation strain fields, i.e., increasing $\Sigma b_i^2$; the dashed curve represents the work of formation for direct homogeneous nucleation of clusters in absence of dislocations, i.e., when $\Sigma b_i^2=0$; (e) the variation of numerically evaluated total energy, $\Delta G_{tot}$, and weighted average burger's vectors, $\Sigma b_i^2$, due to the formation of cluster ``a" in the above box;}
\label{fig:StrainField}
\end{figure*}

The effect of dislocations on the nucleation of clusters is investigated by considering the following form of work of formation:
\begin{align}
W &= 2\pi R\gamma + \pi R^2 (-\Delta f + \Delta G_s) - \Delta G_{sr} + \Delta G_d
\label{Nucl.En}
\end{align}
where $R$ is the cluster radius in terms of number of lattice spacings and
\begin{align}
\gamma = \frac {\int_{Area} {\alpha|\vec{\nabla}c|^2 dr}}{L} 
\label{Surf.En}
\end{align}
is a Cahn-Hilliard type interfacial free energy per unit length of the interface in 2D. $Area$ represents the area of the surface containing the cluster, and $L$ is the circumferential length of a round cluster of radius $R$. Assuming low dislocation density in the system, the interfacial free energy is taken to be solely chemical, neglecting the structural contributions~\cite{Turnbull}.
\begin{align}
\Delta f=f^b-\mu_c^b|_{c^b}(c^b-c^{cl})-f^{cl} 
\label{DrivingForce}
\end{align}
is the bulk driving force for nucleation of a cluster at a given concentration, where superscripts `$b$' and `$cl$' denote the bulk matrix and cluster ``phase'' quantities, respectively.
\begin{align}
\Delta G_s = 2 G_A \delta^2 \frac{K_B}{K_B+G_A} 
\label{StrainEnergy}
\end{align}
represents the strain energy for a coherent nucleus~\cite{hoyt}, where $\delta$ is the misfit strain and $G_A$ and $K_B$ are 2D shear and bulk moduli, respectively, calculated from PFC 2D mode approximation~\cite{greenwood11-2}. 
\begin{align}
\Delta G_{sr} = \eta^2\chi_d E A \ln(R) 
\label{StressRelaxE}
\end{align}
is defined as the stress relaxation term due to segregation of solute into dislocations~\cite{cahn57}, where $A= \frac{G_A \Sigma b_i^2}{4 \pi (1-\nu)}$, $\nu = \frac{E}{2 G_A}-1$, $\eta=\frac{1}{a}\frac{\partial a}{\partial c}$ is the linear expansion coefficient with respect to concentration, $E$ is the 2D Young's modulus~\cite{greenwood11-2}, $\chi_d=(\frac{\partial^2 f}{\partial c^2})^{-1}$, $\Sigma b_i^2$ represents a weighted average of the burger's vectors around the dislocations accompanying the cluster and $a$ is the lattice parameter. The prefactor of the logarithm term, $\eta^2\chi_d E A$, approximates how strain energy is reduced due to solute segregation around a dislocation~\cite{larché85}. 
\begin{align}
\Delta G_d = \zeta A 
\label{Disl.E}
\end{align}
accounts for the increase in the total system energy due to presence of dislocations, where $\zeta$ is a prefactor of order ten giving the average amount of energy per dislocation core~\cite{Hull}. Fig.~\ref{fig:StrainField}(d) plots the evaluation of the above form of work of formation (Eq.~(\ref{Nucl.En})) for cluster ``a" at different mean concentrations up to that of the largest cluster shown in Fig.~\ref{fig:StrainField}(c). The mean concentration of each cluster is estimated within a radius of $R$, defined by radially averaging the radius of the concentration field bound by a threshold of $[c^b+\frac{\sum^N {c-c^b}}{N}]$. The dashed curve in Fig.~\ref{fig:StrainField}(d) represents the work of formation for direct homogeneous nucleation of clusters in absence of dislocations, i.e., $\Sigma b_i^2=0$. The energy barrier for homogeneous nucleation seems to be smaller than that of the dislocation-assisted clustering by a single dislocation, i.e., $\Sigma b_i^2=1$. However, according to the plots shown in Fig.~\ref{fig:StrainField}(d), as Cahn~\cite{cahn57} also pointed out, the barrier for formation of clusters on dislocations can be significantly reduced or even completely eliminated by increasing the magnitude of strain field around the dislocations (i.e. increasing $\Sigma b_i^2$). Notably, the local minimum also shifts to larger nucleus sizes until it vanishes (i.e., work of formation continuously slops down vs. $R$). 

It is noteworthy that, in the absence of quenched-in defects, nucleation of the second phase requires introduction of a thermally-activated noise to produce fluctuations in both density and concentration fields. Assuming dislocations are present in the bulk matrix of a supersaturated quenched alloy, in this study, we demonstrate how elasticity itself can drive the system into a phase transition. The influence of a thermally-activated noise on the transformation kinetics will be investigated in a future study through use of a well-defined noise algorithm. We have, however, observed in our simulations that in the case of a mismatch between the two species, such as in Al-Cu alloys, introducing a Gaussian noise to both density and concentration fields will not have a major impact on the overall path of the transformation. In other words, the phase transformation is mainly driven by the interactions between the elastic fields of the dislocations and the solute atoms.

The total work of formation, $\Delta G_{tot}$, is also estimated numerically by measuring the change in the grand potential within a box engulfing cluster ``a" during its formation and growth in the bulk matrix, i.e., 
\begin{align}
\Delta G_{tot} = \int_V\Omega-\int_V\Omega^b =\nline\int_V {[f-\mu_c c - \mu_n n]} &-\int_V {[f^b -\mu_c^b c^b - \mu_n^b n^b]}
\label{TotalE}
\end{align}
. Here, $\mu_c=\frac{\partial f}{\partial c}$ and $\mu_n=\frac{\partial f}{\partial n}$ are diffusion potentials of concentration and density fields, respectively, and $V$ is the total volume. The above work of formation has contributions from the interfacial energy and driving force for formation of clusters (i.e., $\Delta G_{tot}=\Delta G_{\gamma}-\Delta G_v$), both of which include the elastic effects. Since the above box contains only one cluster, the calculated change in the grand potential accounts for the structural and compositional changes during the formation and growth of only cluster ``a". While the growth of cluster ``a" raises the local free energy, other parts of the system may undergo a process of annihilation and/or shrinkage of sub-critical clusters and their accompanying dislocations leading to an overall decrease in the free energy of the system. As can be seen in Fig.~\ref{fig:StrainField}(e), the total work of formation increases with the growth of cluster ``a" until a maximum value, after which it starts to decrease. Also, as can be seen in this figure, the estimated values of $\Sigma b_i^2$ at various sizes of cluster ``a" closely corresponds to its analytical relationship with the cluster size at the local minima mapped on the energy plots of Fig.~\ref{fig:StrainField}(d). Likewise the work of formation, during formation and growth of cluster ``a", the value of $\Sigma b_i^2$ reaches a maximum at the critical size of the cluster. 

According to our data, cluster ``a" continuously grows in presence of dislocations implying that, at each sub-critical cluster size, the system is sitting at a local energy minimum. Since cluster ``a" at each sub-critical size is in a local equilibrium with the matrix we call it a metastable precursor to the cluster ``a" with a critical size. This is analogous to previous PFC studies of crystals solidification which show that metastable amorphous precursors emerge first due to their lower nucleation barrier than that of a crystalline solid~\cite{toth11,tegze09}. In our case, the nucleation barrier is lowered by the effect of locally straining a sub-critical cluster (as a result of local accumulation of dislocations burger's vectors, as illustrated in Fig.~\ref{fig:Clustering}(a)), making it thermodynamically favorable for the cluster to receive more solute atoms from the matrix and grow in size. 

\begin{figure}[htbp]
\resizebox{3.4in}{!}{\includegraphics{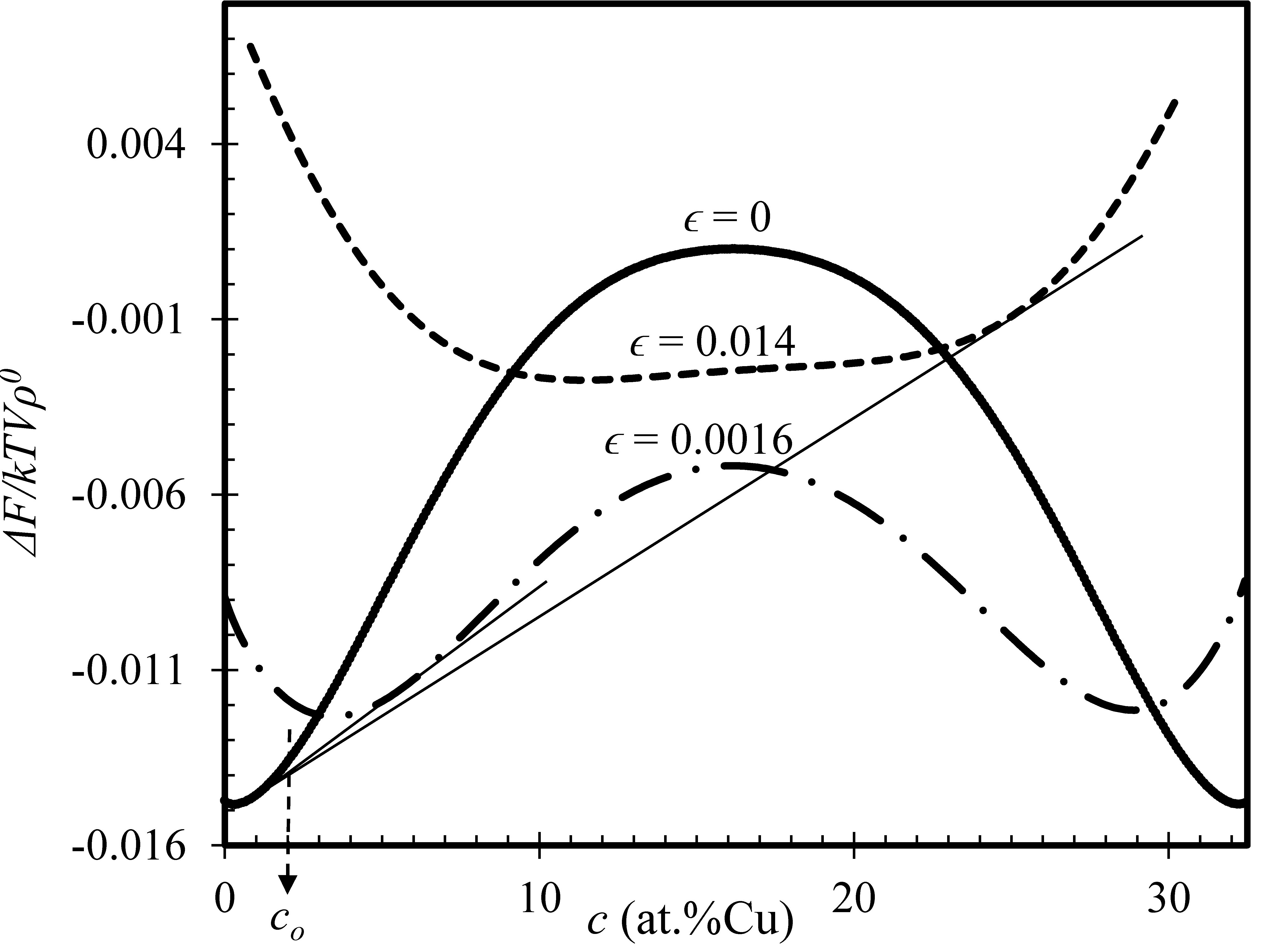}}
\caption{(colour online) Common tangent construction using mean-field free energy curves of unstrained (solid curve) and strained solid phases (dashed curves).}
\label{fig:StrainELand}
\end{figure}

\subsection{Metastable phase coexistence}
The metastable coexistence between a sub-critical cluster ``a" and the matrix at the quench/aging temperature is elucidated by evaluating the mean field free energy of a system comprising an unstrained matrix phase and strained solid phases with different magnitudes of distortion, i.e., a uniform strain. The free energy-concentration curve of a strained solid phase, at a given temperature, can be achieved by calculating the peaks of correlation kernel $\hat{C}_{2j}^{ii}$, at locations slightly off those of the equilibrium density peaks, $k_j$, for a square structure. The introduced amount of strain is defined in Fourier space by
\begin{align}
\epsilon=|k-k_j|/k_j
\label{FourierStrain}
\end{align}
, where index $j$ denotes one family of planes in reciprocal space. As can be inferred from Fig.~\ref{fig:StrainELand}, increasing the amount of strain from 0.0016 to 0.014 (corresponding approximately to the average strain within the cluster ``a" shown in Fig.~\ref{fig:StrainField}(b) and (c), respectively), raises the free energy in the strained solid. The free energy wells also shift to different concentrations of solute. Such a configuration admits a common tangent between the free energy curves of the unstrained matrix (e.g. the solid curve) and the distorted ones (e.g. dashed curves)~\cite{larché85}, leading to a (metastable) multiphase coexistence with a lower free energy (as demonstrated in Fig.~\ref{fig:StrainELand}). In other words, at each level of local strain, there is a thermodynamic driving force for a transformation from a single-phase structure of a strained matrix to a phase-coexistence between a strained cluster and an unstrained matrix. On the other hand, despite the fact that the above transformation is thermodynamically favorable, the configuration of energy plots in Fig.~\ref{fig:StrainELand} implies that the driving force for nucleation is lower for the strained cluster (i.e., using the definition of $\Delta f$ in Eq.~(\ref{Nucl.En})). However, since the energy curves in this figure are derived from a mean-field PFC approximation, the illustrated phase-coexistence does not carry the effect of interfacial energy and only includes a mean-field sense of the misfit strain. These factors have a significant impact on the thermodynamics of phase-coexistence at cluster sizes smaller than that of the critical nucleus. In fact, the previously described stress relaxation term in the definition of work of formation (Eq.~(\ref{StressRelaxE})), $\Delta G_{sr}$, overcompensates for the effect of reduced driving force for formation of strained clusters.  

\subsection{Clustering mechanism}
Based on our PFC simulations, we propose the following mechanism of clustering: (1) Stress relaxation by segregation of solute atoms into highly-strained areas in the matrix, such as around dislocations, (2) strain-aided nucleation of sub-critical clusters at concentrations higher than that of the matrix and (3) subsequent growth and enrichment of sub-critical clusters into overcritical sizes, only if a sufficient strain field is preserved, to overcome the nucleation barrier. The above mechanism is consistent with the experimentally observed formation and enrichment of highly-strained coherent GP-zones in quenched-aged dilute Al-Cu alloys~\cite{biswas11}, proposed as the initial step before precipitation of the semi-coherent and incoherent equilibrium $\theta$-phases~\cite{christian}. GP-zones in dilute binary Al alloys are normally known as coherent/semi-coherent particles often with a crystal structure and composition similar to those of the final equilibrium precipitate~\cite{christian,hoyt}. Our clusters possess the same chemical composition and lattice parameter as those of the equilibrium theta-phase pre-set by the relevant peaks in our correlation functions. Thus, they would represent an early-stage evolution of the so-called GP-zones. An investigation on the transformation of GP-zones into the subsequent metastable and equilibrium precipitates will be followed in a future study in 3D with more complex crystal structures. We expect to observe a gradual loss of coherency as GP-zones grow in size, as dictated by the energy arguments. We also note that we expect our results to hold qualitatively in 3D, since the same type of elastic effects are expected to appear around the dislocations regardless of their dimension and any possible partial splitting of dislocations around the clusters.

\section{Summary}
In summary, we showed that the alloy phase field crystal model of ref.~\cite{greenwood11-2} which stabilizes different crystal structures can be used to simulate and analyze the mechanisms of clustering phenomenon in bulk lattice of quenched/aged alloys. In accordance with the existing experimental observations, our simulations suggests that quenched-in defects, such as dislocations, significantly lower the energy barrier for nucleation of clusters. Furthermore, analysis of overall system energy and local energy changes reveal that the formation and growth of sub-critical clusters are thermodynamically favorable in conjunction with quenched-in mobile dislocations. Consistent with existing experiments, our simulations shed significant light on the elusive energetic mechanism of the growth and enrichment of early clusters which are the precursors of bulk precipitation. 

\begin{acknowledgements}
We acknowledge the financial support received from National Science and Engineering Research Council of Canada
(NSERC), Ontario Ministry of Research and Innovation (Early Researcher Award Program) and the Clumeq High Performance Centre.
\end{acknowledgements}

\bibliography{references}

\begin{thebibliography}{30}%
\makeatletter
\providecommand \@ifxundefined [1]{%
 \@ifx{#1\undefined}
}%
\providecommand \@ifnum [1]{%
 \ifnum #1\expandafter \@firstoftwo
 \else \expandafter \@secondoftwo
 \fi
}%
\providecommand \@ifx [1]{%
 \ifx #1\expandafter \@firstoftwo
 \else \expandafter \@secondoftwo
 \fi
}%
\providecommand \natexlab [1]{#1}%
\providecommand \enquote  [1]{``#1''}%
\providecommand \bibnamefont  [1]{#1}%
\providecommand \bibfnamefont [1]{#1}%
\providecommand \citenamefont [1]{#1}%
\providecommand \href@noop [0]{\@secondoftwo}%
\providecommand \href [0]{\begingroup \@sanitize@url \@href}%
\providecommand \@href[1]{\@@startlink{#1}\@@href}%
\providecommand \@@href[1]{\endgroup#1\@@endlink}%
\providecommand \@sanitize@url [0]{\catcode `\\12\catcode `\$12\catcode
  `\&12\catcode `\#12\catcode `\^12\catcode `\_12\catcode `\%12\relax}%
\providecommand \@@startlink[1]{}%
\providecommand \@@endlink[0]{}%
\providecommand \url  [0]{\begingroup\@sanitize@url \@url }%
\providecommand \@url [1]{\endgroup\@href {#1}{\urlprefix }}%
\providecommand \urlprefix  [0]{URL }%
\providecommand \Eprint [0]{\href }%
\providecommand \doibase [0]{http://dx.doi.org/}%
\providecommand \selectlanguage [0]{\@gobble}%
\providecommand \bibinfo  [0]{\@secondoftwo}%
\providecommand \bibfield  [0]{\@secondoftwo}%
\providecommand \translation [1]{[#1]}%
\providecommand \BibitemOpen [0]{}%
\providecommand \bibitemStop [0]{}%
\providecommand \bibitemNoStop [0]{.\EOS\space}%
\providecommand \EOS [0]{\spacefactor3000\relax}%
\providecommand \BibitemShut  [1]{\csname bibitem#1\endcsname}%
\let\auto@bib@innerbib\@empty
\bibitem [{\citenamefont {Christian}(2002)}]{christian}%
  \BibitemOpen
  \bibfield  {author} {\bibinfo {author} {\bibfnamefont {J.~W.}\ \bibnamefont
  {Christian}},\ }\enquote {\bibinfo {title} {The theory of transformations in
  metals and alloys},}\ \ (\bibinfo  {publisher} {Pergamon, Oxford, UK},\
  \bibinfo {year} {2002})\ \bibinfo {edition} {2nd}\ ed.\BibitemShut {Stop}%
\bibitem [{\citenamefont {Marceau}\ \emph
  {et~al.}(2010{\natexlab{a}})\citenamefont {Marceau}, \citenamefont {Sha},
  \citenamefont {Lumley},\ and\ \citenamefont {Ringer}}]{marceau10}%
  \BibitemOpen
  \bibfield  {author} {\bibinfo {author} {\bibfnamefont {R.}~\bibnamefont
  {Marceau}}, \bibinfo {author} {\bibfnamefont {G.}~\bibnamefont {Sha}},
  \bibinfo {author} {\bibfnamefont {R.}~\bibnamefont {Lumley}}, \ and\ \bibinfo
  {author} {\bibfnamefont {S.}~\bibnamefont {Ringer}},\ }\href {\doibase
  10.1016/j.actamat.2009.11.021} {\bibfield  {journal} {\bibinfo  {journal}
  {Acta Mater.}\ }\textbf {\bibinfo {volume} {58}},\ \bibinfo {pages} {1795 }
  (\bibinfo {year} {2010}{\natexlab{a}})}\BibitemShut {NoStop}%
\bibitem [{\citenamefont {Marceau}\ \emph
  {et~al.}(2010{\natexlab{b}})\citenamefont {Marceau}, \citenamefont {Sha},
  \citenamefont {Ferragut}, \citenamefont {Dupasquier},\ and\ \citenamefont
  {Ringer}}]{marceau10-2}%
  \BibitemOpen
  \bibfield  {author} {\bibinfo {author} {\bibfnamefont {R.}~\bibnamefont
  {Marceau}}, \bibinfo {author} {\bibfnamefont {G.}~\bibnamefont {Sha}},
  \bibinfo {author} {\bibfnamefont {R.}~\bibnamefont {Ferragut}}, \bibinfo
  {author} {\bibfnamefont {A.}~\bibnamefont {Dupasquier}}, \ and\ \bibinfo
  {author} {\bibfnamefont {S.}~\bibnamefont {Ringer}},\ }\href {\doibase
  10.1016/j.actamat.2010.05.020} {\bibfield  {journal} {\bibinfo  {journal}
  {Acta Materialia}\ }\textbf {\bibinfo {volume} {58}},\ \bibinfo {pages} {4923
  } (\bibinfo {year} {2010}{\natexlab{b}})}\BibitemShut {NoStop}%
\bibitem [{\citenamefont {Nutyen}(1967)}]{nutyen67}%
  \BibitemOpen
  \bibfield  {author} {\bibinfo {author} {\bibfnamefont {J.~B.~M.}\
  \bibnamefont {Nutyen}},\ }\href@noop {} {\bibfield  {journal} {\bibinfo
  {journal} {Acta Metall.}\ }\textbf {\bibinfo {volume} {15}},\ \bibinfo
  {pages} {1765} (\bibinfo {year} {1967})}\BibitemShut {NoStop}%
\bibitem [{\citenamefont {Ozawa}\ and\ \citenamefont {Kimura}(1970)}]{ozawa70}%
  \BibitemOpen
  \bibfield  {author} {\bibinfo {author} {\bibfnamefont {E.}~\bibnamefont
  {Ozawa}}\ and\ \bibinfo {author} {\bibfnamefont {H.}~\bibnamefont {Kimura}},\
  }\href {\doibase 10.1016/0001-6160(70)90055-6} {\bibfield  {journal}
  {\bibinfo  {journal} {Acta Metall.}\ }\textbf {\bibinfo {volume} {18}},\
  \bibinfo {pages} {995 } (\bibinfo {year} {1970})}\BibitemShut {NoStop}%
\bibitem [{\citenamefont {Sato}\ \emph {et~al.}(2004)\citenamefont {Sato},
  \citenamefont {Hirose},\ and\ \citenamefont {Hirosawa}}]{sato04}%
  \BibitemOpen
  \bibfield  {author} {\bibinfo {author} {\bibfnamefont {T.}~\bibnamefont
  {Sato}}, \bibinfo {author} {\bibfnamefont {K.}~\bibnamefont {Hirose}}, \ and\
  \bibinfo {author} {\bibfnamefont {S.}~\bibnamefont {Hirosawa}},\ }\href@noop
  {} {\bibfield  {journal} {\bibinfo  {journal} {Mater. Forum}\ }\textbf
  {\bibinfo {volume} {28}},\ \bibinfo {pages} {956} (\bibinfo {year}
  {2004})}\BibitemShut {NoStop}%
\bibitem [{\citenamefont {Esmaeili}\ \emph {et~al.}(2007)\citenamefont
  {Esmaeili}, \citenamefont {Vaumousse}, \citenamefont {Zandbergen},
  \citenamefont {Poole}, \citenamefont {Cerezo},\ and\ \citenamefont
  {Lloyd}}]{esmaeili07}%
  \BibitemOpen
  \bibfield  {author} {\bibinfo {author} {\bibfnamefont {S.}~\bibnamefont
  {Esmaeili}}, \bibinfo {author} {\bibfnamefont {D.}~\bibnamefont {Vaumousse}},
  \bibinfo {author} {\bibfnamefont {M.~W.}\ \bibnamefont {Zandbergen}},
  \bibinfo {author} {\bibfnamefont {W.~J.}\ \bibnamefont {Poole}}, \bibinfo
  {author} {\bibfnamefont {A.}~\bibnamefont {Cerezo}}, \ and\ \bibinfo {author}
  {\bibfnamefont {D.~J.}\ \bibnamefont {Lloyd}},\ }\href@noop {} {\bibfield
  {journal} {\bibinfo  {journal} {Philos. Mag.}\ }\textbf {\bibinfo {volume}
  {87}},\ \bibinfo {pages} {3797} (\bibinfo {year} {2007})}\BibitemShut
  {NoStop}%
\bibitem [{\citenamefont {Biswas}\ \emph {et~al.}(2011)\citenamefont {Biswas},
  \citenamefont {Siegel}, \citenamefont {Wolverton},\ and\ \citenamefont
  {Seidman}}]{biswas11}%
  \BibitemOpen
  \bibfield  {author} {\bibinfo {author} {\bibfnamefont {A.}~\bibnamefont
  {Biswas}}, \bibinfo {author} {\bibfnamefont {D.~J.}\ \bibnamefont {Siegel}},
  \bibinfo {author} {\bibfnamefont {C.}~\bibnamefont {Wolverton}}, \ and\
  \bibinfo {author} {\bibfnamefont {D.~N.}\ \bibnamefont {Seidman}},\ }\href
  {\doibase 10.1016/j.actamat.2011.06.036} {\bibfield  {journal} {\bibinfo
  {journal} {Acta Materialia}\ }\textbf {\bibinfo {volume} {59}},\ \bibinfo
  {pages} {6187 } (\bibinfo {year} {2011})}\BibitemShut {NoStop}%
\bibitem [{\citenamefont {Somoza}\ \emph {et~al.}(2002)\citenamefont {Somoza},
  \citenamefont {Petkov}, \citenamefont {Lynn},\ and\ \citenamefont
  {Dupasquier}}]{somoza02}%
  \BibitemOpen
  \bibfield  {author} {\bibinfo {author} {\bibfnamefont {A.}~\bibnamefont
  {Somoza}}, \bibinfo {author} {\bibfnamefont {M.~P.}\ \bibnamefont {Petkov}},
  \bibinfo {author} {\bibfnamefont {K.~G.}\ \bibnamefont {Lynn}}, \ and\
  \bibinfo {author} {\bibfnamefont {A.}~\bibnamefont {Dupasquier}},\ }\href
  {\doibase 10.1103/PhysRevB.65.094107} {\bibfield  {journal} {\bibinfo
  {journal} {Phys. Rev. B}\ }\textbf {\bibinfo {volume} {65}},\ \bibinfo
  {pages} {094107} (\bibinfo {year} {2002})}\BibitemShut {NoStop}%
\bibitem [{\citenamefont {Somoza}\ and\ \citenamefont
  {Dupasquier}(2010)}]{somoza10}%
  \BibitemOpen
  \bibfield  {author} {\bibinfo {author} {\bibfnamefont {A.}~\bibnamefont
  {Somoza}}\ and\ \bibinfo {author} {\bibnamefont {Dupasquier}},\ }in\
  \href@noop {} {\emph {\bibinfo {booktitle} {Fundamentals of Aluminium
  Metallurgy: Production, Processing and Applications}}},\ \bibinfo {editor}
  {edited by\ \bibinfo {editor} {\bibfnamefont {R.}~\bibnamefont {Lumley}}}\
  (\bibinfo  {publisher} {Woodhead Publishing},\ \bibinfo {year} {2010})\
  \bibinfo {edition} {1st}\ ed.,\ pp.\ \bibinfo {pages} {386--421}\BibitemShut
  {NoStop}%
\bibitem [{\citenamefont {Jaatinen}\ \emph {et~al.}(2009)\citenamefont
  {Jaatinen}, \citenamefont {Achim}, \citenamefont {Elder},\ and\ \citenamefont
  {Ala-Nissila}}]{jaatinen09}%
  \BibitemOpen
  \bibfield  {author} {\bibinfo {author} {\bibfnamefont {A.}~\bibnamefont
  {Jaatinen}}, \bibinfo {author} {\bibfnamefont {C.~V.}\ \bibnamefont {Achim}},
  \bibinfo {author} {\bibfnamefont {K.~R.}\ \bibnamefont {Elder}}, \ and\
  \bibinfo {author} {\bibfnamefont {T.}~\bibnamefont {Ala-Nissila}},\ }\href
  {\doibase 10.1103/PhysRevE.80.031602} {\bibfield  {journal} {\bibinfo
  {journal} {Phys. Rev. E}\ }\textbf {\bibinfo {volume} {80}},\ \bibinfo
  {pages} {031602} (\bibinfo {year} {2009})}\BibitemShut {NoStop}%
\bibitem [{\citenamefont {Elder}\ and\ \citenamefont {Grant}(2004)}]{elder04}%
  \BibitemOpen
  \bibfield  {author} {\bibinfo {author} {\bibfnamefont {K.~R.}\ \bibnamefont
  {Elder}}\ and\ \bibinfo {author} {\bibfnamefont {M.}~\bibnamefont {Grant}},\
  }\href {\doibase 10.1103/PhysRevE.70.051605} {\bibfield  {journal} {\bibinfo
  {journal} {Phys. Rev. E}\ }\textbf {\bibinfo {volume} {70}},\ \bibinfo
  {pages} {051605} (\bibinfo {year} {2004})}\BibitemShut {NoStop}%
\bibitem [{\citenamefont {Elder}\ \emph {et~al.}(2007)\citenamefont {Elder},
  \citenamefont {Provatas}, \citenamefont {Berry}, \citenamefont {Stefanovic},\
  and\ \citenamefont {Grant}}]{elder07}%
  \BibitemOpen
  \bibfield  {author} {\bibinfo {author} {\bibfnamefont {K.~R.}\ \bibnamefont
  {Elder}}, \bibinfo {author} {\bibfnamefont {N.}~\bibnamefont {Provatas}},
  \bibinfo {author} {\bibfnamefont {J.}~\bibnamefont {Berry}}, \bibinfo
  {author} {\bibfnamefont {P.}~\bibnamefont {Stefanovic}}, \ and\ \bibinfo
  {author} {\bibfnamefont {M.}~\bibnamefont {Grant}},\ }\href {\doibase
  10.1103/PhysRevB.75.064107} {\bibfield  {journal} {\bibinfo  {journal} {Phys.
  Rev. B}\ }\textbf {\bibinfo {volume} {75}},\ \bibinfo {pages} {064107}
  (\bibinfo {year} {2007})}\BibitemShut {NoStop}%
\bibitem [{\citenamefont {Wu}\ \emph {et~al.}(2010)\citenamefont {Wu},
  \citenamefont {Adland},\ and\ \citenamefont {Karma}}]{wu10}%
  \BibitemOpen
  \bibfield  {author} {\bibinfo {author} {\bibfnamefont {K.-A.}\ \bibnamefont
  {Wu}}, \bibinfo {author} {\bibfnamefont {A.}~\bibnamefont {Adland}}, \ and\
  \bibinfo {author} {\bibfnamefont {A.}~\bibnamefont {Karma}},\ }\href
  {\doibase 10.1103/PhysRevE.81.061601} {\bibfield  {journal} {\bibinfo
  {journal} {Phys. Rev. E}\ }\textbf {\bibinfo {volume} {81}},\ \bibinfo
  {pages} {061601} (\bibinfo {year} {2010})}\BibitemShut {NoStop}%
\bibitem [{\citenamefont {Greenwood}\ \emph {et~al.}(2010)\citenamefont
  {Greenwood}, \citenamefont {Provatas},\ and\ \citenamefont
  {Rottler}}]{greenwood10}%
  \BibitemOpen
  \bibfield  {author} {\bibinfo {author} {\bibfnamefont {M.}~\bibnamefont
  {Greenwood}}, \bibinfo {author} {\bibfnamefont {N.}~\bibnamefont {Provatas}},
  \ and\ \bibinfo {author} {\bibfnamefont {J.}~\bibnamefont {Rottler}},\ }\href
  {\doibase 10.1103/PhysRevLett.105.045702} {\bibfield  {journal} {\bibinfo
  {journal} {Phys. Rev. Lett.}\ }\textbf {\bibinfo {volume} {105}},\ \bibinfo
  {pages} {045702} (\bibinfo {year} {2010})}\BibitemShut {NoStop}%
\bibitem [{\citenamefont {Greenwood}\ \emph
  {et~al.}(2011{\natexlab{a}})\citenamefont {Greenwood}, \citenamefont
  {Rottler},\ and\ \citenamefont {Provatas}}]{greenwood11}%
  \BibitemOpen
  \bibfield  {author} {\bibinfo {author} {\bibfnamefont {M.}~\bibnamefont
  {Greenwood}}, \bibinfo {author} {\bibfnamefont {J.}~\bibnamefont {Rottler}},
  \ and\ \bibinfo {author} {\bibfnamefont {N.}~\bibnamefont {Provatas}},\
  }\href {\doibase 10.1103/PhysRevE.83.031601} {\bibfield  {journal} {\bibinfo
  {journal} {Phys. Rev. E}\ }\textbf {\bibinfo {volume} {83}},\ \bibinfo
  {pages} {031601} (\bibinfo {year} {2011}{\natexlab{a}})}\BibitemShut
  {NoStop}%
\bibitem [{\citenamefont {Greenwood}\ \emph
  {et~al.}(2011{\natexlab{b}})\citenamefont {Greenwood}, \citenamefont
  {Ofori-Opoku}, \citenamefont {Rottler},\ and\ \citenamefont
  {Provatas}}]{greenwood11-2}%
  \BibitemOpen
  \bibfield  {author} {\bibinfo {author} {\bibfnamefont {M.}~\bibnamefont
  {Greenwood}}, \bibinfo {author} {\bibfnamefont {N.}~\bibnamefont
  {Ofori-Opoku}}, \bibinfo {author} {\bibfnamefont {J.}~\bibnamefont
  {Rottler}}, \ and\ \bibinfo {author} {\bibfnamefont {N.}~\bibnamefont
  {Provatas}},\ }\href {\doibase 10.1103/PhysRevB.84.064104} {\bibfield
  {journal} {\bibinfo  {journal} {Phys. Rev. B}\ }\textbf {\bibinfo {volume}
  {84}},\ \bibinfo {pages} {064104} (\bibinfo {year}
  {2011}{\natexlab{b}})}\BibitemShut {NoStop}%
\bibitem [{\citenamefont {Archer}(2005)}]{archer05}%
  \BibitemOpen
  \bibfield  {author} {\bibinfo {author} {\bibfnamefont {A.~J.}\ \bibnamefont
  {Archer}},\ }\href {http://stacks.iop.org/0953-8984/17/i=10/a=001} {\bibfield
   {journal} {\bibinfo  {journal} {Journal of Physics: Condensed Matter}\
  }\textbf {\bibinfo {volume} {17}},\ \bibinfo {pages} {1405} (\bibinfo {year}
  {2005})}\BibitemShut {NoStop}%
\bibitem [{\citenamefont {Baker}\ and\ \citenamefont {Okamoto}(1993)}]{baker}%
  \BibitemOpen
  \bibfield  {author} {\bibinfo {author} {\bibfnamefont {H.}~\bibnamefont
  {Baker}}\ and\ \bibinfo {author} {\bibfnamefont {H.}~\bibnamefont
  {Okamoto}},\ }\enquote {\bibinfo {title} {Alloy phase diagrams},}\ \
  (\bibinfo  {publisher} {ASM International, Materials Park, Ohio},\ \bibinfo
  {year} {1993})\ \bibinfo {edition} {1st}\ ed.\BibitemShut {Stop}%
\bibitem [{\citenamefont {Desorbo}\ \emph {et~al.}(1958)\citenamefont
  {Desorbo}, \citenamefont {Treaftis},\ and\ \citenamefont
  {Turnbull}}]{desorbo58}%
  \BibitemOpen
  \bibfield  {author} {\bibinfo {author} {\bibfnamefont {W.}~\bibnamefont
  {Desorbo}}, \bibinfo {author} {\bibfnamefont {H.}~\bibnamefont {Treaftis}}, \
  and\ \bibinfo {author} {\bibfnamefont {D.}~\bibnamefont {Turnbull}},\ }\href
  {\doibase 10.1016/0001-6160(58)90019-1} {\bibfield  {journal} {\bibinfo
  {journal} {Acta Metall.}\ }\textbf {\bibinfo {volume} {6}},\ \bibinfo {pages}
  {401 } (\bibinfo {year} {1958})}\BibitemShut {NoStop}%
\bibitem [{\citenamefont {Ozawa}\ and\ \citenamefont {Kimura}(1971)}]{ozawa71}%
  \BibitemOpen
  \bibfield  {author} {\bibinfo {author} {\bibfnamefont {E.}~\bibnamefont
  {Ozawa}}\ and\ \bibinfo {author} {\bibfnamefont {H.}~\bibnamefont {Kimura}},\
  }\href {\doibase 10.1016/0025-5416(71)90100-5} {\bibfield  {journal}
  {\bibinfo  {journal} {Materials Science and Engineering}\ }\textbf {\bibinfo
  {volume} {8}},\ \bibinfo {pages} {327 } (\bibinfo {year} {1971})}\BibitemShut
  {NoStop}%
\bibitem [{\citenamefont {Leonard}\ and\ \citenamefont
  {Haataja}(2005)}]{leonard05}%
  \BibitemOpen
  \bibfield  {author} {\bibinfo {author} {\bibfnamefont {F.}~\bibnamefont
  {Leonard}}\ and\ \bibinfo {author} {\bibfnamefont {M.}~\bibnamefont
  {Haataja}},\ }\href {\doibase 10.1063/1.1922578} {\bibfield  {journal}
  {\bibinfo  {journal} {Applied Physics Letters}\ }\textbf {\bibinfo {volume}
  {86}},\ \bibinfo {pages} {181909} (\bibinfo {year} {2005})}\BibitemShut
  {NoStop}%
\bibitem [{\citenamefont {Muralidharan}\ and\ \citenamefont
  {Haataja}(2010)}]{muralidharan10}%
  \BibitemOpen
  \bibfield  {author} {\bibinfo {author} {\bibfnamefont {S.}~\bibnamefont
  {Muralidharan}}\ and\ \bibinfo {author} {\bibfnamefont {M.}~\bibnamefont
  {Haataja}},\ }\href {\doibase 10.1103/PhysRevLett.105.126101} {\bibfield
  {journal} {\bibinfo  {journal} {Phys. Rev. Lett.}\ }\textbf {\bibinfo
  {volume} {105}},\ \bibinfo {pages} {126101} (\bibinfo {year}
  {2010})}\BibitemShut {NoStop}%
\bibitem [{\citenamefont {Turnbull}(1955)}]{Turnbull}%
  \BibitemOpen
  \bibfield  {author} {\bibinfo {author} {\bibfnamefont {D.}~\bibnamefont
  {Turnbull}},\ }\enquote {\bibinfo {title} {Impurities and imperfections},}\ \
  (\bibinfo  {publisher} {American Society of Metals, Ohio},\ \bibinfo {year}
  {1955})\ p.\ \bibinfo {pages} {121}\BibitemShut {NoStop}%
\bibitem [{\citenamefont {Hoyt}(2010)}]{hoyt}%
  \BibitemOpen
  \bibfield  {author} {\bibinfo {author} {\bibfnamefont {J.}~\bibnamefont
  {Hoyt}},\ }\enquote {\bibinfo {title} {Phase transformations},}\ \ (\bibinfo
  {publisher} {Titles on Demand, Ontario, Canada},\ \bibinfo {year} {2010})\
  pp.\ \bibinfo {pages} {87--102},\ \bibinfo {edition} {1st}\ ed.\BibitemShut
  {Stop}%
\bibitem [{\citenamefont {Cahn}(1957)}]{cahn57}%
  \BibitemOpen
  \bibfield  {author} {\bibinfo {author} {\bibfnamefont {J.~W.}\ \bibnamefont
  {Cahn}},\ }\href {\doibase 10.1016/0001-6160(57)90021-4} {\bibfield
  {journal} {\bibinfo  {journal} {Acta Metallurgica}\ }\textbf {\bibinfo
  {volume} {5}},\ \bibinfo {pages} {169 } (\bibinfo {year} {1957})}\BibitemShut
  {NoStop}%
\bibitem [{\citenamefont {Larche}\ and\ \citenamefont
  {Cahn}(1985)}]{larché85}%
  \BibitemOpen
  \bibfield  {author} {\bibinfo {author} {\bibfnamefont {F.}~\bibnamefont
  {Larche}}\ and\ \bibinfo {author} {\bibfnamefont {J.}~\bibnamefont {Cahn}},\
  }\href {\doibase 10.1016/0001-6160(85)90077-X} {\bibfield  {journal}
  {\bibinfo  {journal} {Acta Metallurgica}\ }\textbf {\bibinfo {volume} {33}},\
  \bibinfo {pages} {331 } (\bibinfo {year} {1985})}\BibitemShut {NoStop}%
\bibitem [{\citenamefont {Hull}(1975)}]{Hull}%
  \BibitemOpen
  \bibfield  {author} {\bibinfo {author} {\bibfnamefont {D.}~\bibnamefont
  {Hull}},\ }\enquote {\bibinfo {title} {Introduction to dislocations},}\ \
  (\bibinfo  {publisher} {Pergamon, Oxford, UK},\ \bibinfo {year} {1975})\ pp.\
  \bibinfo {pages} {90--93},\ \bibinfo {edition} {2nd}\ ed.\BibitemShut {Stop}%
\bibitem [{\citenamefont {T\'oth}\ \emph {et~al.}(2011)\citenamefont {T\'oth},
  \citenamefont {Pusztai}, \citenamefont {Tegze}, \citenamefont {T\'oth},\ and\
  \citenamefont {Gr\'an\'asy}}]{toth11}%
  \BibitemOpen
  \bibfield  {author} {\bibinfo {author} {\bibfnamefont {G.~I.}\ \bibnamefont
  {T\'oth}}, \bibinfo {author} {\bibfnamefont {T.}~\bibnamefont {Pusztai}},
  \bibinfo {author} {\bibfnamefont {G.}~\bibnamefont {Tegze}}, \bibinfo
  {author} {\bibfnamefont {G.}~\bibnamefont {T\'oth}}, \ and\ \bibinfo {author}
  {\bibfnamefont {L.}~\bibnamefont {Gr\'an\'asy}},\ }\href {\doibase
  10.1103/PhysRevLett.107.175702} {\bibfield  {journal} {\bibinfo  {journal}
  {Phys. Rev. Lett.}\ }\textbf {\bibinfo {volume} {107}},\ \bibinfo {pages}
  {175702} (\bibinfo {year} {2011})}\BibitemShut {NoStop}%
\bibitem [{\citenamefont {Tegze}\ \emph {et~al.}(2009)\citenamefont {Tegze},
  \citenamefont {Gr\'an\'asy}, \citenamefont {T\'oth}, \citenamefont
  {Podmaniczky}, \citenamefont {Jaatinen}, \citenamefont {Ala-Nissila},\ and\
  \citenamefont {Pusztai}}]{tegze09}%
  \BibitemOpen
  \bibfield  {author} {\bibinfo {author} {\bibfnamefont {G.}~\bibnamefont
  {Tegze}}, \bibinfo {author} {\bibfnamefont {L.}~\bibnamefont {Gr\'an\'asy}},
  \bibinfo {author} {\bibfnamefont {G.~I.}\ \bibnamefont {T\'oth}}, \bibinfo
  {author} {\bibfnamefont {F.}~\bibnamefont {Podmaniczky}}, \bibinfo {author}
  {\bibfnamefont {A.}~\bibnamefont {Jaatinen}}, \bibinfo {author}
  {\bibfnamefont {T.}~\bibnamefont {Ala-Nissila}}, \ and\ \bibinfo {author}
  {\bibfnamefont {T.}~\bibnamefont {Pusztai}},\ }\href {\doibase
  10.1103/PhysRevLett.103.035702} {\bibfield  {journal} {\bibinfo  {journal}
  {Phys. Rev. Lett.}\ }\textbf {\bibinfo {volume} {103}},\ \bibinfo {pages}
  {035702} (\bibinfo {year} {2009})}\BibitemShut {NoStop}%
\end{thebibliography}%

\end{document}